\documentclass[pra,aps,twocolumn,nopacs,superscriptaddress,nofootinbib]{revtex4}
\usepackage{graphicx}
\usepackage{dcolumn}
\usepackage{bm}
\usepackage{amsmath}
\usepackage{epsfig}
\usepackage{color}

\def\ii{{\rm i}} \def\ee{{\rm e}}  \def\jb{{\bf j}}
\def\Eb{{\bf E}} \def\Rb{{\bf R}}

 \def\varepsv{\vec{\varepsilon}}

\def\kb{{\bf k}}

\def\EF{{E_{\rm F}}}  \def\vF{{v_{\rm F}}}  \def\vF2{{v_{\rm F}^2}}

\def\thetav{\vec{\theta}}

\begin{document}

\title{Quantum Effects in the Nonlinear Response of Graphene Plasmons}

\author{Joel~D.~Cox}
\email{joel.cox@icfo.es}
\affiliation{ICFO-Institut de Ciencies Fotoniques, The Barcelona Institute of Science and Technology, 08860 Castelldefels (Barcelona), Spain}
\author{Iv\'an~Silveiro}
\affiliation{ICFO-Institut de Ciencies Fotoniques, The Barcelona Institute of Science and Technology, 08860 Castelldefels (Barcelona), Spain}
\author{F.~Javier~Garc\'{\i}a~de~Abajo}
\affiliation{ICFO-Institut de Ciencies Fotoniques, The Barcelona Institute of Science and Technology, 08860 Castelldefels (Barcelona), Spain}
\affiliation{ICREA-Instituci\'o Catalana de Recerca i Estudis Avan\c{c}ats, Passeig Llu\'{\i}s Companys 23, 08010 Barcelona, Spain}

\begin{abstract}
The ability of graphene to support long-lived, electrically tunable plasmons that interact strongly with light, combined with its highly nonlinear optical response, has generated great expectations for application of the atomically-thin material to nanophotonic devices. These expectations are mainly reinforced by classical analyses performed using the response derived from extended graphene, neglecting finite-size and nonlocal effects that become important when the carbon layer is structured on the nanometer scale in actual device designs. Here we show that finite-size effects produce large contributions that increase the nonlinear response of nanostructured graphene to significantly higher levels than those predicted by classical theories. We base our analysis on a quantum-mechanical description of graphene using tight-binding electronic states combined with the random-phase approximation. While classical and quantum descriptions agree well for the linear response when either the plasmon energy is below the Fermi energy or the size of the structure exceeds a few tens of nanometers, this is not always the case for the nonlinear response, and in particular, third-order Kerr-type nonlinearities are generally underestimated by the classical theory. Our results reveal the complex quantum nature of the optical response in nanostructured graphene, while further supporting the exceptional potential of this material for nonlinear nanophotonic devices.
\end{abstract}
%\date{\today}
\maketitle

\section{Introduction}

Graphene, the two-dimensional monolayer of carbon atoms arranged in a honeycomb lattice, has proven to be an ideal material platform for numerous photonic and opto-electronic applications \cite{BSH10, BL12}. Owing to its gapless, linear charge-carrier dispersion relation \cite{W1947, CGP09}, graphene is capable of relatively strong, broadband coupling with light \cite{NBG08,MSW08}, and offers a large electrically-tunable optical response \cite{LYU11}. Valence and conduction electrons in this material present a uniform velocity that clearly emphasizes their anharmonic response to external fields, which has stimulated considerable interest in the graphene nonlinear optical properties \cite{M07_2,HHM10,DD10,WZY11,ZVB12,GPM12,KKG13,HDP13,GG14,PBS14,YTB14,PC14,NA14,CVS14,CVS15,M15_2}. Further motivation to study optical nonlinearities in monolayer graphene is provided by the availability of interband optical transitions within a continuous range of low photon energies, accompanied by a high electrical mobility \cite{NGM04, NGM05}. The large third-order susceptibilities corresponding to four-wave mixing \cite{HHM10, GPM12}, the optical Kerr effect \cite{WZY11, ZVB12}, and third-harmonic generation \cite{KKG13, HDP13} (THG) that have been measured in graphene are indeed attributed to these electronic properties. However, the superiority of graphene with respect to conventional nonlinear optical semiconductors for ultrafast all-optical switching is not yet fully established \cite{K14}, although this atomically thin material has been shown to be an excellent broadband saturable absorber for laser mode locking \cite{BL12}.

The intrinsically-high nonlinear optical response of graphene can be further enhanced by plasmonic excitations supported by the carbon layer when it is electrically doped. Unlike {\it traditional} plasmons in noble metals, whose resonance frequencies are determined by geometric and intrinsic properties, graphene plasmons can be actively tuned by modifying the doping density \cite{JGH11,FAB11,paper196,FRA12,YLC12,YLL12,paper212,BJS13,paper230,YLZ13,FLZ14,GPN12,paper235}. Because of its superior electrical conductivity, plasmons in graphene are also longer-lived and couple more efficiently to light than their noble-metal counterparts. Additionally, they exhibit extreme subwavelength confinement \cite{JBS09,paper176}, so that the excitation of propagating plasmons in extended graphene often relies on near-field coupling to satisfy energy-momentum conservation \cite{FAB11,paper196,FRA12}. This constraint is relaxed in nanostructured graphene, enabling optical excitation of localized plasmons from free space \cite{YXL12,paper212,BJS13,YLZ13}. The strong near electric fields generated by plasmons in noble metals have been widely exploited to enhance nonlinear optical processes \cite{DN07,KZ12,BBM15}, and thus, even larger enhancements are expected from graphene plasmons, as recently predicted by several theoretical studies \cite{M11,paper226,G13_2,SSS13_2,NBN13,YTB14,paper247,paper250,SNS15,JC15,CYJ15,paper259}.

For extended graphene, the nonlinear response is commonly described using a classical nonlinear conductivity derived from the Boltzmann transport equation (BTE) \cite{M07_2,PBS14} assuming that intraband transitions dominate the optical response (\textit{i.e.}, for photon energies $\hbar\omega$ roughly below the Fermi energy $\EF$). At higher energies, a quantum-mechanical (QM) treatment is required to account for the linear and nonlinear response arising from interband electron transitions \cite{CVS14,CVS15,M15_2}. This approach has been used to predict a large nonlinear optical susceptibility in extended graphene at low doping (\textit{i.e.}, without plasmonic enhancement), for which reasonable agreement with experimental observations has been reported \cite{HHM10}.

To describe plasmon-enhanced nonlinear optical processes in nanostructured graphene, a finite graphene structure is typically assigned a local nonlinear conductivity, and its response simulated within the framework of classical electrodynamics \cite{paper226,SNS15,CYJ15,paper259}. However, this classical approach neglects nonlocal, finite-size, and atomistic (\textit{e.g.}, edge termination) effects, all of which have been found to play crucial roles in describing both the linear \cite{paper183,CWJ14,WCJ15} and nonlinear \cite{paper247,paper250} optical response of graphene structures with small ($\sim10$\,nm) features.
Here we reveal strong finite-size and atomistic effects in the nonlinear response of graphene nanoribbons and nanoislands, predicted to take place from a realistic, QM description of these structures, beyond classical theory, which can produce dramatically different results for structure sizes up to a few tens of nanometers. The discrepancy between classical and QM descriptions is particularly large for the Kerr effect/nonlinear absorption (\textit{i.e.}, the complex third-order susceptibility oscillating at the fundamental frequency, which from here onwards will be referred to simply as the ``Kerr nonlinearity'') where classical theory underestimates the strength of the third-order response by several orders of magnitude even for $>20\,$nm structures. This is at odds with the conclusions previously drawn by examining the linear response regime, where the plasmon frequencies and strengths were found to be similar within classical and QM approaches either when the structures were a few tens of nanometers in size or when they did not possess zigzag edges. The QM effects here reported for such large sizes represent a rather unusual scenario in plasmonics, while they support the use of doped nanographene structures as plasmon-driven nonlinear enhancers, where they perform much better than previously estimated from the study of extended graphene.

%%%%%%%%%%%%%%%%%%%%%%%%%%%%%%%%%%%%%%%%%%%%%%%%%%%%%%%%%%%%%%%%%%%%%%%%%%

\section{Results and discussion}

\begin{figure*}[t]
\includegraphics[width=1\textwidth]{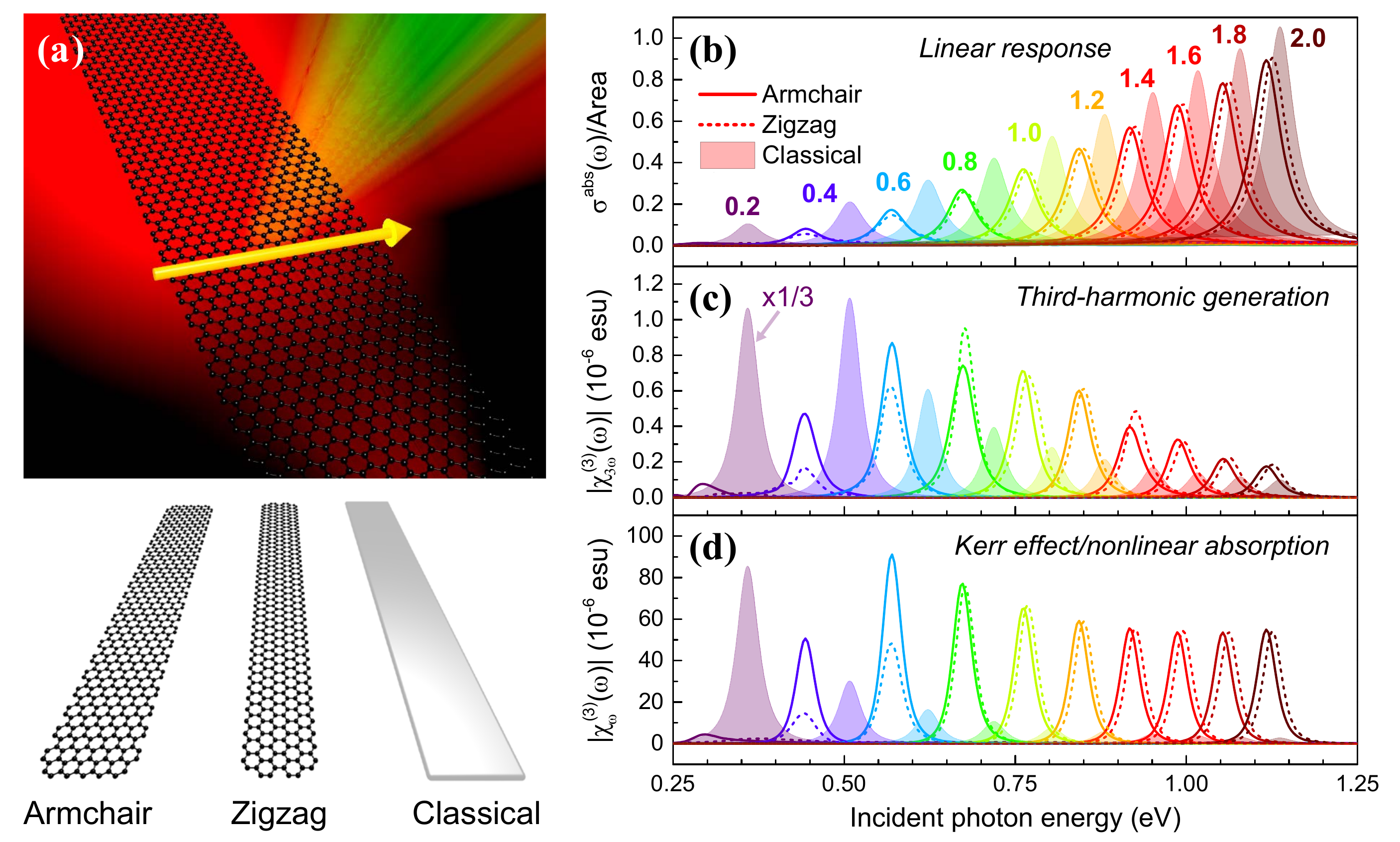}
\caption{{\bf Linear and nonlinear spectral response of graphene nanoribbons.} {\bf (a)} We consider cw incident light linearly polarized across the ribbons (top) and present results derived from a quantum-mechanical (QM) model (tight-binding+random-phase approximation) for structures with either armchair or zigzag edges (bottom, left and center), compared with classical electromagnetic simulations (local conductivity) for a homogeneous planar sheet (bottom, right). {\bf (b-d)} Linear absorption cross-section (b), and third-order susceptibilities for THG (c, where the spectrum for $E_F=0.2$\,eV has been scaled by 1/3) and the Kerr nonlinearity (d), as obtained from the QM model for armchair (solid curves) and zigzag (dashed curves) nanoribbons, compared with classical electrodynamic simulations (filled curves). Different Fermi energies (color-coded numerical values in (b), eV) are considered, taking the ribbon width as $\approx10\,$nm and the damping $\hbar\tau^{-1}=50$\,meV in all cases.}
\label{rib_spectra}
\end{figure*}

\subsection{Graphene nanoribbons}

 We study graphene nanoribbons illuminated with light polarized perpendicular to their direction of translational symmetry. As second-order processes are forbidden in this geometry \cite{B08_3}, we concentrate on first- and third-order phenomena. In Fig.\ \ref{rib_spectra} we examine the incident-frequency dependence of the linear and nonlinear polarizabilities for ribbons roughly 10\,nm in width. In particular, we present atomistic QM simulations for ribbons of armchair (carbon-to-carbon width 9.97\,nm) and zigzag (width 9.81\,nm) edges, compared with classical theory (width 10\,nm). Details of the calculation procedures are given in the Appendix. The linear response (Fig.\ \ref{rib_spectra}b) is characterized by a prominent plasmon feature, for which the classical calculations are in reasonable agreement with the quantum model for both types of edges, provided that the Fermi energy $\EF$ is larger than the plasmon energy. The only discrepancy in this regime is a small systematic plasmon redshift and broadening in the QM model relative to the classical simulations. In contrast, when the plasmon energy is above $\EF$, the quantum model predicts a strong plasmon attenuation and the shift becomes dramatic.

Remarkably, similar agreement between classical and quantum models is observed in THG at high doping (Fig.\ \ref{rib_spectra}c), which deteriorates as the Fermi energy is lowered, again producing an overestimate of the classical response compared with the QM simulations. This is unlike the Kerr nonlinearity (Fig.\ \ref{rib_spectra}d), which we find to be substantially smaller in the classical description, a result that could originate in the inability of this model to account for the simultaneous plasmonic enhancement of the frequency-degenerate input and output waves. The values of $\chi^{(3)}(\omega)$ predicted in the QM description for graphene nanoribbons are on the order of $10^{-5}$\,esu ($10^{-12}$\,m$^2$/V$^{2}$), to be compared with those measured in gold, ranging from $10^{-7}$--$10^{-13}$\,esu ($10^{-14}$--$10^{-20}$\,m$^2$/V$^{2}$) \cite{BSD14} (values for the nonlinear refractive index/nonlinear absorption coefficient are provided in Supporting Information, SI). We also note that these values depend strongly on the inelastic scattering decay time, $\tau$, for which we have used a highly conservative value $\hbar\tau^{-1}=50$\,meV, or $\tau\simeq13$\,fs, corresponding to moderate values of the DC mobility less than 660\,cm$^2/($V\,s$)$, as estimated from the Drude model \cite{AM1976} for the Fermi energies under consideration.

\begin{figure*}[t]
\includegraphics[width=1\textwidth]{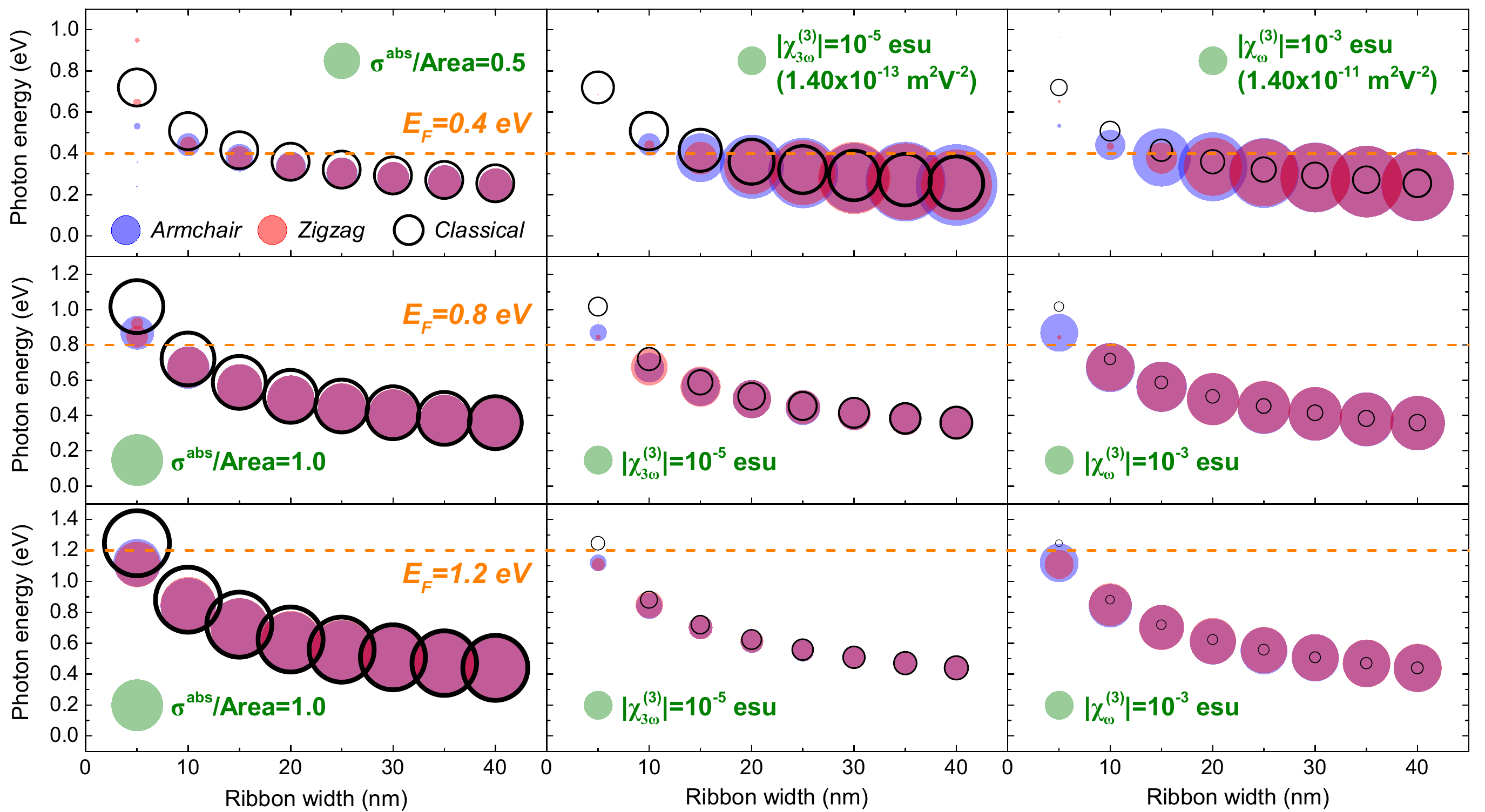}
\caption{{\bf Evolution of the nonlinear ribbon response with size and doping level.} The symbols indicate the peak maxima for the linear absorption cross-section (left column), as well as the nonlinear polarizabilities corresponding to THG (central column) and the Kerr nonlinearity (right column), calculated for the structures shown in Fig.\ \ref{rib_spectra}a. Results are presented for QM calculations of armchair (solid blue circles) and zigzag (solid red circles) ribbons, compared with classical simulations (open circles). Each of the three rows corresponds to a different Fermi energy ($\EF=0.4, 0.8,$ and $1.2$\,eV from top to bottom), and in all cases we have used $\hbar\tau^{-1}=20$\,meV ($\tau\simeq33$\,fs). The peak areas are proportional to the symbol areas (see upper legends in each plot).}
\label{rib_bubbles}
\end{figure*}

An overview of the size and doping dependence of the linear and nonlinear ribbon polarizabilities is presented in Fig.\ \ref{rib_bubbles}. Again, the linear absorption of armchair and zigzag ribbons in the quantum description displays very similar plasmon frequencies and strengths when the plasmon energy is below $\EF$ (\textit{i.e.}, for widths $\sim5-15\,$nm, depending on the actual value of $\EF$ under consideration), and they in turn agree with the classical simulations. For larger plasmon energies (smaller sizes) quantum effects become important, reflecting in particular a strong coupling to electronic edge states in zigzag ribbons \cite{paper249}. We extract similar conclusions regarding the ability of classical theory to describe THG (Fig.\ \ref{rib_bubbles}), in good qualitative agreement with QM simulations, except at low doping levels and small ribbon widths. However, classical theory underestimates the polarizability associated with the Kerr nonlinearity by roughly an order of magnitude, even for ribbon widths as large as 40\,nm.

%%%%%%%%%%%%%%%%%%%%%%%%%%%%%%%%%%%%%%%%%%%%%%%%%%%%%%%%%%%%%%%%%%%%%%%%%%

\subsection{Graphene nanoislands}

\begin{figure*}[t]
\includegraphics[width=1\textwidth]{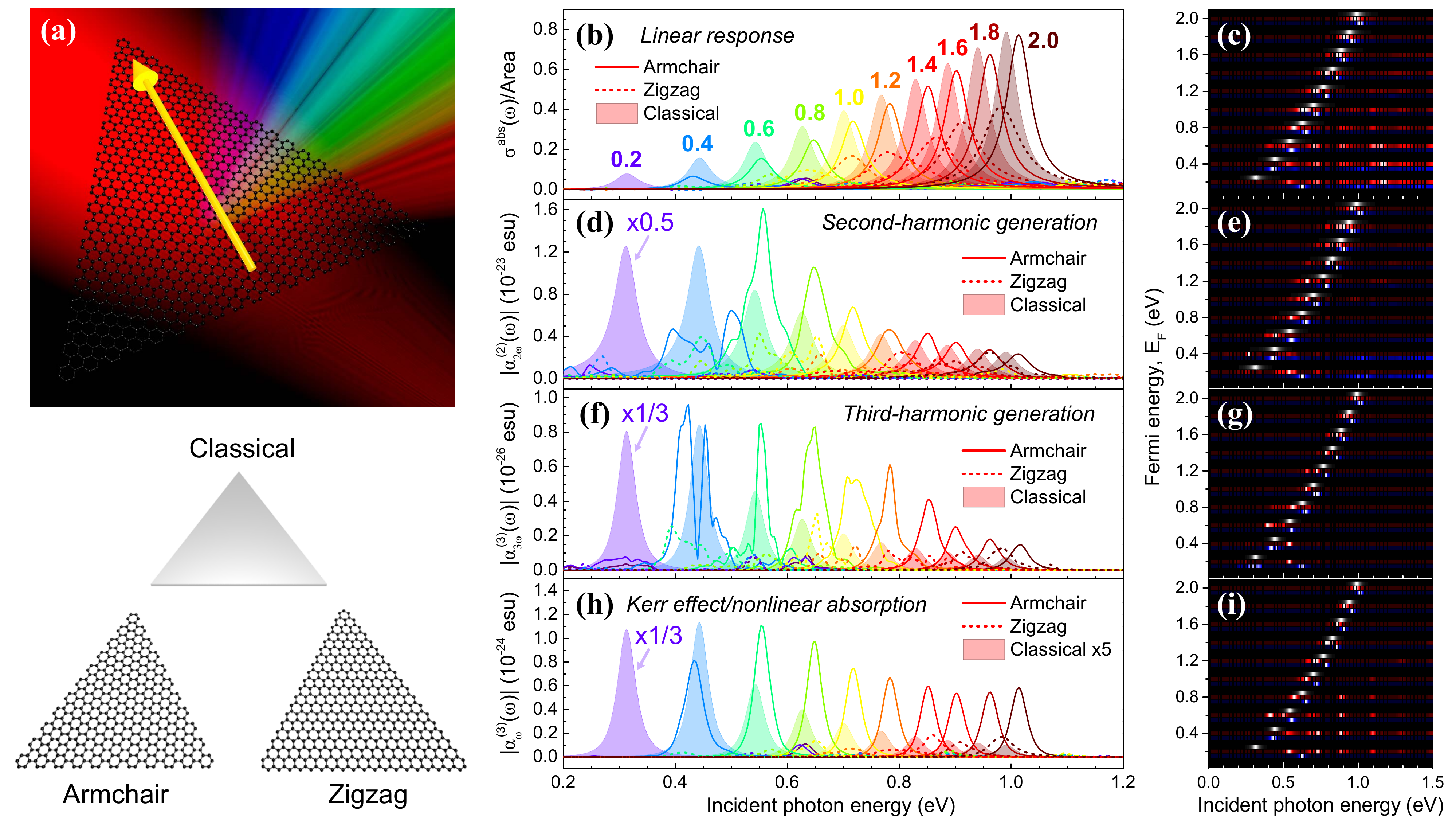}
\caption{{\bf Linear and nonlinear spectral response of graphene nanotriangles.} {\bf (a)} We consider light polarization as indicated in the upper plot and present QM simulations for structures with armchair or zigzag edges (bottom), compared with classical electromagnetic results for a homogeneous equilateral triangle (center). {\bf (b-i)} Linear absorption cross-section (b,c), normalized to the nanotriangle area, and susceptibilities for second-harmonic generation (SHG) (d,e), THG (f,g), and the Kerr nonlinearity (h,i), as obtained from the QM model for armchair (solid curves) and zigzag (dashed curves) nanotriangles, compared with classical electrodynamic simulations (filled curves). Different Fermi energies (color-coded numerical values in (b), eV) are considered. In all cases, the triangle sides are $\approx10\,$nm and we take $\hbar\tau^{-1}=50$\,meV. The spectra for $E_F=0.2$\,eV have been scaled by the indicated factors. The contour plots on the right (c,e,g,i) show the corresponding normalized absorption spectra combining the results from QM simulations of armchair (blue) and zigzag (red) islands, alongside classical calculations (grey).}
\label{tri_spectra}
\end{figure*}

In Fig.\ \ref{tri_spectra}, we examine the spectral response of the linear and nonlinear polarizability for equilateral triangular nanographenes roughly 10\,nm in side length and illuminated by light polarized in a direction perpendicular to one of the triangle sides (see Fig.\ \ref{tri_spectra} top). Despite the centrosymmetry of the graphene crystal structure, even-ordered, dipolar nonlinear processes are enabled by the symmetry-breaking of finite structures lacking inversion symmetry along the direction of the induced nonlinear dipole moment. The chosen triangular shape allows us to study islands with approximately equal dimensions, but containing exclusively either armchair- or zigzag-terminated edges (see Fig.\ \ref{tri_spectra}a, bottom). The linear absorption cross-sections, normalized to the nanoisland areas, are presented in Fig.\ \ref{tri_spectra}b,c, where, for lower doping, we find dramatic differences between atomistic and classical results, and also strong discrepancies among the two types of edge terminations considered in the atomistic simulations. With increasing Fermi level, the numerous plasmon resonances that appear due to the finite-size and nonlocal effects captured by the QM description tend to coalesce towards the prominent dipolar plasmon mode predicted by the classical simulations. We also note that the spectral features of the zigzag island are typically smaller in magnitude than those of the armchair island, which is attributed to the presence of zero-energy electronic states confined to the zigzag edges \cite{paper214,CWJ14}.

Our results employing the classical and quantum descriptions of the second-order nonlinear polarizability corresponding to second-harmonic generation (SHG) (Fig.\ \ref{tri_spectra}d,e) reveal that the classical theory provides good qualitative agreement with QM simulations, in particular for armchair-edged islands, when plasmon resonances are at energies below $E_F$. The classical description relies on the second-order conductivity for SHG in extended graphene, for which the leading contribution is nonlocal due to photon momentum transfer \cite{paper259}. Conversely, the atomistic simulations are not limited by any such assumed symmetry, and also account for the responses at both the fundamental and second harmonic frequencies, which can significantly enhance SHG at low doping levels \cite{paper247}. This is similar to surface-dipole-induced SHG, which is known to be generally stronger than bulk quadrupole \cite{DD10,BBM15}. Nevertheless, our findings suggest that the nonlocal electric field enhancement associated with a plasmon resonance, which is well described by our classical model when $E_F$ is larger than the resonance frequency, provides the leading contribution to SHG in nanoislands.

The third-order nonlinear polarizabilities associated with THG and the Kerr nonlinearity are presented in Fig.\ \ref{tri_spectra}f,g and Fig.\ \ref{tri_spectra}h,i, respectively, for the triangular nanoisland. We find fairly good agreement between the THG polarizabilities calculated using the classical and quantum approaches at high doping, while for the Kerr nonlinearity the classical treatment underestimates the polarizability by one order of magnitude compared with the quantum approach, except for very low doping levels. Conversely, the Kerr nonlinearity involves the mixing of three waves at the fundamental frequency to produce a fourth wave, also oscillating at the fundamental frequency. At the dominant plasmon mode, all four waves are then simultaneously enhanced by the near electric field of the plasmon. This enhancement at the {\it output} frequency is not taken into account in the classical description, resulting in a significantly weaker Kerr response.

\begin{figure*}[t]
\includegraphics[width=1\textwidth]{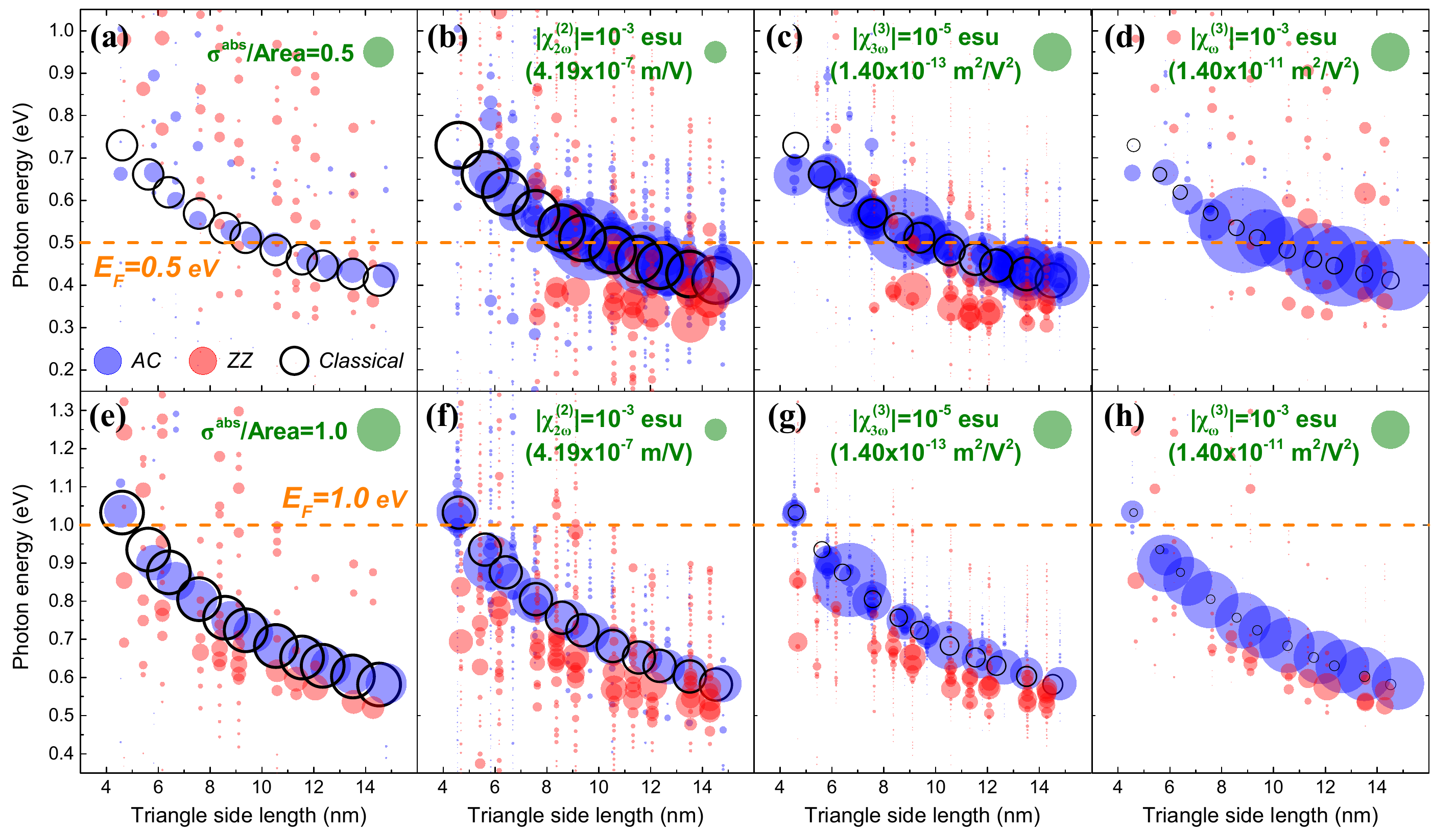}
\caption{{\bf Evolution of the nonlinear nanotriangle response with size and doping level.} The symbols indicate the peak maxima for the linear absorption cross-section (a,e), as well as the nonlinear susceptibilities corresponding to SHG (b,f), THG (c,g), and the Kerr nonlinearity (d,h), calculated for the structures shown in Fig.\ \ref{tri_spectra}a. Results are presented for QM calculations of armchair (solid blue circles) and zigzag (solid red circles) triangles, compared with classical simulations (open circles). The upper (lower) row corresponds to a Fermi energy $\EF=0.5\,$eV ($\EF=1\,$eV), and in all cases we use $\hbar\tau^{-1}=20$\,meV. The peak maxima are proportional to the symbol areas, where the scales are indicated by the green legends.}
\label{tri_bubbles}
\end{figure*}

Figure\ \ref{tri_bubbles} presents an overview of our results for nanotriangles. As expected, the linear response shows that the dipolar plasmon resonances predicted in the classical treatment generally agree more closely with the atomistic simulations for armchair-edged nanoislands. For SHG and THG, the nonlinear polarizabilities obtained from the classical treatment are in good agreement with the atomistic simulations for the armchair islands, presumably for similar reasons as in the linear response, while the quantum simulations predict a much larger nonlinear response for the Kerr nonlinearity, which we attribute to the simultaneous enhancement at the fundamental and mixed frequencies --an effect that is not captured in the classical description.

%%%%%%%%%%%%%%%%%%%%%%%%%%%%%%%%%%%%%%%%%%%%%%%%%%%%%%%%%%%%%%%%%%%%%%%%%%

\section{Conclusions}

We have demonstrated that the nonlinear optical response of nanostructured graphene is strongly influenced by nonlocal and finite-size effects, which cannot easily be ignored in structures with dimensions of $<20$\,nm, and quite possibly in even larger systems. The classical electrodynamic description fails to capture these effects, which we have found to be particularly important when dealing with the optical Kerr nonlinearity. Interestingly, THG in nanoribbons is very well described by the classical approach, which also provides reasonable agreement with quantum-mechanical results for graphene nanoislands, though better agreement is obtained for predominantly armchair-edged islands. Qualitatively similar agreement between quantum-mechanical and classical descriptions is also found for SHG in nanoislands lacking inversion symmetry, indicating that the second-order nonlocal {\it bulk} nonlinear conductivity, combined with the dipolar electric field associated with a plasmon mode, is sufficient to describe SHG in such structures. Barring edge-effects in nanoislands, it is then apparent that SHG and THG are reasonably well described by the classical electrodynamic simulations when plasmon energies are much less than the chemical potential gap $2\EF$, yet the optical Kerr nonlinearity is lacking significant contributions from the nonlocal response at the fundamental frequency, for which four waves are involved; in this case, a more fundamental quantum-mechanical description predicts a much higher nonlinear response than classical theory, thus providing further support for the use of highly doped graphene as an excellent nonlinear element in nanophotonic devices.

%%%%%%%%%%%%%%%%%%%%%%%%%%%%%%%%%%%%%%%%%%%%%%%%%%%%%%%%%%%%%%%%%%%%%%%%%%

\appendix

\begin{widetext}

\section{Classical electrodynamical simulations}

Our classical calculations are based upon the linear and nonlinear conductivities for extended graphene $\sigma^{(1)}_\omega$, $\sigma^{(2)}_{2\omega}$, $\sigma^{(3)}_{3\omega}$, and $\sigma^{(3)}_\omega$, which describe linear, second-harmonic, third-harmonic, and Kerr-type processes, respectively. As explained in more detail in the SI, we adopt an eigenmode expansion of the graphene optical response \cite{paper228} and concentrate on a spectral region dominated by a single plasmon, whose in-plane electric field is $\varepsv_1(\Rb)$. Assuming a local dielectric function $\epsilon=1+4\pi\ii\sigma^{(1)}_{\omega}(\omega)/\omega t$, where $t$ the thickness of the graphene layer (here we take $t=0.5$\,nm, which is found to be well converged with respect to the $t\rightarrow0$ limit), we calculate the field $\varepsv_1(\Rb)$ for equilateral triangles using a finite-element method, and normalize it in such a way that $\int_S d^2\Rb |\varepsv_1(\Rb)|^2=D^2$, where $S$ denotes the 2D region occupied by the graphene structure and $D$ is a characteristic length (\textit{e.g.}, the triangle side length). We now define $\eta^{(n)}_{s\omega}=\ii\sigma^{(n)}_{s\omega}/s\omega D$, and consider, without loss of generality, a plasmon polarized along a symmetry direction $x$. Then, the linear polarizability is given by
\begin{equation}
\alpha^{(1)}_{\omega}(\omega)=\frac{\eta^{(1)}_\omega\xi_1^2 D^3}{1-\eta^{(1)}_\omega(\omega)/\eta_1},
\end{equation}
where \[\xi_1=-\left(1/D^2\right)\int_S d^2\Rb\,\varepsilon_{1,x}\] and $\eta_1$ is a plasmon eigenvalue \cite{paper228}, which is determined from the calculated linear extinction spectrum. Similarly, the nonlinear optical polarizabilities associated with SHG, THG, and the Kerr nonlinearity are given as
\begin{align}
\alpha^{(2)}_{2\omega}(\omega)&=\frac{\eta^{(2)}_{2\omega}\xi_1^2\zeta^{(2)}_{2\omega}D^2}{\left[1-\eta^{(1)}_\omega/\eta_1\right]^2}, \\
\alpha^{(3)}_{3\omega}(\omega)&=\frac{\eta^{(3)}_{3\omega}\xi_1^3\zeta^{(3)}_{3\omega}D^3}{\left[1-\eta^{(1)}_\omega/\eta_1\right]^3}, \\
\alpha^{(3)}_{\omega}(\omega)&=\frac{\eta^{(3)}_{\omega}\xi_1^3\zeta^{(3)}_\omega D^3}{\left|1-\eta^{(1)}_\omega/\eta_1\right|^2\left[1-\eta^{(1)}_\omega/\eta_1\right]},
\end{align}
respectively, where
\[\zeta^{(2)}_{2\omega}=\left(1/D\right)\int_S d^2\Rb \left[\varepsilon_{1,x} \left(\partial_x\varepsilon_{1,x}+5\partial_y\varepsilon_{1,y}/3\right)+\varepsilon_{1,y}\left(\partial_y\varepsilon_{1,x}/3-\partial_x\varepsilon_{1,y}\right)\right],\] \[\zeta^{(3)}_{3\omega}=-\left(1/D^2\right)\int_S d^2\Rb \left(\varepsv_{1} \cdot \varepsv_1\right) \varepsilon_{1,x},\] and \[\zeta^{(3)}_{\omega}=-\left(1/3D^2\right)\int_S d^2\Rb \left(2|\varepsv_1|^2 \varepsilon_{1,x}+\varepsv_1\cdot\varepsv_1\varepsilon^*_{1,x}\right)\] are dimensionless coefficients, for which numerical values are provided in Table\ \ref{Table1}. Nonlinear susceptibilities are estimated by normalizing the polarizabilities to the nanostructure volume, assuming an effective thickness of $0.33$\,nm for the graphene layer (i.e., the interplane distance in graphite) \cite{HHM10,paper247}.

For one-dimensional graphene nanoribbons, we use the boundary-element method \cite{paper040} to calculate the in-plane electric field $\varepsv_1(\Rb)$, taking $D$ as the ribbon width. Following the same procedure as for finite structures, we recover Eqs.\ (1--4) with the left-hand side of each equation divided by the (infinite) ribbon length, so that $\alpha^{(n)}_{s\omega}$ instead denotes the polarizability per unit length along the ribbon for an $n^{\text{th}}$-order process at harmonic $s$ of the illumination frequency.

\begin{table*}
\begin{center}
\begin{tabular}{lccccc}
\hline
\hline
  & $\eta_1$ & $\xi_1$ & $\zeta^{(2)}_{2\omega}$ & $\zeta^{(3)}_{3\omega}$ & $\zeta^{(3)}_{\omega}$ \\
\hline
Triangle &-0.0933 &0.541 &-1.90 &1.57 &1.57\\
\hline
Ribbon &-0.0709 &0.951 &N/A &1.46 &1.46\\
\hline
\hline
\end{tabular}
\end{center}
\caption{
Numerical Values for the parameters used to calculate the classical polarizabilities.}
\label{Table1}
\end{table*}

\section{Nonlinear conductivities in extended graphene}

The linear and nonlinear surface currents in graphene are obtained \textit{via} iterative solution of the Boltzmann transport equation \cite{PBS14},
\begin{equation}\label{boltz}
\frac{\partial f_\kb(\Rb, t)}{\partial t}-\frac{e}{\hbar}\textbf{E}\cdot\nabla_\kb f_\kb(\Rb, t)\pm v_{\rm F} \frac{\kb}{k}\cdot\nabla_\Rb f_\kb(\Rb, t)=-\frac{1}{\tau}\left[ f_\kb(\Rb, t) -  f_\kb^0\right],
\end{equation}
where $f_\kb^0$ is the Fermi-Dirac distribution, to which the system relaxes at the phenomenological scattering rate $\tau^{-1}$, and we consider the interaction with an ac field $\Eb(t)=\Eb_{\omega}(\ee^{-\ii\omega t}+\ee^{\ii\omega t})$ that is polarized parallel to the graphene plane with polarization across the ribbon (\textit{i.e.}, the $x$-$y$ plane). At first-order, in the long-wavelength limit, the linear surface current is $\jb^{(1)}_\omega(\omega)=\sigma^{(1)}_\omega(\omega)\Eb_\omega$, with the conductivity
\begin{equation}
\sigma^{(1)}_{\omega}(\omega)=\frac{\ii e^2 E_F}{\pi\hbar^2(\omega+\ii\tau^{-1})},
\end{equation}
which coincides with the Drude model, also resulting from the local limit of the random-phase approximation (RPA) after neglecting interband transitions. 

The dominant contribution to the second-harmonic current $\jb^{(2)}_{2\omega}(\omega)$ arises from the lowest-order expansion in the photon momentum (\textit{i.e.}, a nonlocal contribution originating in the $\nabla_\Rb$ term of Eq.\ (\ref{boltz})), which leads to \cite{paper259}
\begin{equation}
j^{(2)}_{2\omega,i}(\omega)=\sigma^{(2)}_{2\omega}(\omega)\sum_{jkl}\left(\frac{5}{3}\delta_{ij}\delta_{kl}-\delta_{ik}\delta_{jl}+\frac{1}{3}\delta_{il}\delta_{jk}\right)E_{\omega,j}\partial_k E_{\omega,l},
\label{bbbb}
\end{equation}
where the scalar part of the second-harmonic conductivity tensor is given by
\begin{equation}
\sigma^{(2)}_{2\omega}(\omega)=\frac{3\ii e^3v_F^2}{8\pi\hbar^2(\omega+\ii\tau^{-1})^3}
\end{equation}
and the $i$ subindex in Eq.\ (\ref{bbbb}) refers to the direction of the second-harmonic current.

The leading contribution to the third-harmonic and Kerr-type nonlinear currents is local (\textit{i.e.}, we can neglect the spatial variations in $f_\kb(\Rb,t)$). For THG, the nonlinear surface current is $\jb^{(3)}_{3\omega}(\omega)=\sigma^{(3)}_{3\omega}(\omega)\left(\Eb_\omega\cdot\Eb_\omega\right)\Eb_\omega$, where the conductivity reads
\begin{equation}
\sigma^{(3)}_{3\omega}(\omega)=\frac{3\ii e^4v_F^2}{4\pi\hbar^2 E_F(\omega+\ii\tau^{-1})(2\omega+\ii\tau^{-1})(3\omega+\ii\tau^{-1})},
\end{equation}
while for the Kerr nonlinearity we have $\jb^{(3)}_{\omega}(\omega)=\sigma^{(3)}_{3\omega}(\omega)\left[2|\Eb_\omega|^2\Eb_\omega+(\Eb_\omega\cdot\Eb_\omega)\Eb_\omega^*\right]/3$, with
\begin{equation}
\sigma^{(3)}_{\omega}(\omega)=\frac{9\ii e^4\vF2}{4\pi\hbar^2\EF(\omega+\ii\tau^{-1})(-\omega+\ii\tau^{-1})(2\omega+\ii\tau^{-1})}.
\end{equation}
Note that the above expressions for the linear and nonlinear conductivities are obtained in the zero-temperature limit, which should be valid for $T=300$\,K provided $\hbar\omega\gg k_{\rm B}T$ (see SI for a comparison of results obtained at $T=0$ and 300 K).

\section{Quantum-mechanical simulations}

We employ a tight-binding Hamiltonian $H_{TB}$ to describe the $\pi$-band electronic structure of graphene nanoislands and nanoribbons, where each carbon atom is represented by a single $p$ orbital oriented perpendicular to the graphene plane, and a hopping energy of 2.8\,eV connects nearest-neighbor carbon sites. To simulate the optical response of these systems we solve the equation of motion for the single-electron density matrix,
\begin{equation}\label{rho_eom}
\frac{\partial \rho}{\partial t} = -\frac{\ii}{\hbar} \left[ H_{TB}-e\phi, \rho \right] - \frac{1}{2\tau} \big( \rho - \rho^{0} \big),
\end{equation}
where $\phi$ is the self-consistent electric potential. The last term in Eq.\ (\ref{rho_eom}) describes inelastic electron scattering with a phenomenological decay time $\tau$, taking $\rho^{0}$ as the $t\rightarrow -\infty$ equilibrium density matrix. The polarizability spectra of Figs.\ \ref{rib_spectra} and \ref{tri_spectra} are obtained with $\hbar\tau^{-1}=50$\,meV, while we take $\hbar\tau^{-1}=20$\,meV for the polarizability maps of Figs.\ \ref{rib_bubbles} and \ref{tri_bubbles}. These damping rates correspond to moderate values of the DC mobility up to 1646\,\color{black}cm$^2/($V\,s$)$, as estimated from the Drude model \cite{AM1976} for the Fermi energies under consideration.

For a graphene nanoribbon consisting of $N$ unit cells with period $b$ along its direction of translational symmetry, electronic Bloch states are constructed as
\begin{equation}\label{bloch}
|j,k\rangle=\frac{1}{\sqrt{N}}\sum_{l,m}a_{jl,k}\ee^{\ii kmb}|l,m\rangle,
\end{equation}
where $j$ denotes the band index, $k$ is the in-plane Bloch wave vector along the ribbon, and $l$ labels carbon sites $\textbf{R}_l$ within each unit cell $m$. We then write the density matrix as
\begin{equation}\label{rho_def}
\rho=\sum_{jj',kk'}\rho_{jj',kk'}|j,k\rangle\langle j',k'|=\sum_{jj',kk',ll'}\rho_{jj',kk'}a_{jl,k}a^*_{j'l',k'}|l,k\rangle\langle l',k'|,
\end{equation}
where we have defined
\begin{equation}\label{lk_states}
|l,k\rangle=\frac{1}{\sqrt{N}}\sum_m\ee^{\ii kmb}|l,m\rangle.
\end{equation}
Eqs.\ (\ref{bloch}) and (\ref{lk_states}) yield the relations $\rho_{ll',kk'}=\sum_{jj'}\rho_{jj',kk'}a_{jl,k}a^*_{j'l',k'}$ and $\rho_{jj',kk'}=\sum_{ll'}\rho_{ll',kk'}a^*_{jl,k}a_{j'l',k'}$, which connect site and state representations, allowing us to express the equation of motion in the latter as
\begin{align}\label{rhojk_eom}
\frac{\partial \rho_{jj',kk'}}{\partial t}&=-\ii\left(\varepsilon_{j,k}-\varepsilon_{j',k'}\right)\rho_{jj',kk'}+\frac{\ii e}{\hbar}\sum_{j''k''}\left(\phi_{jj'',kk''}\rho_{j''j',k''k'}-\rho_{jj'',kk''}\phi_{j''j',k''k'}\right) \\
&\quad-\frac{1}{2\tau}\big(\rho_{jj',kk'}-\rho^0_{jj',kk'}\big) \nonumber,
\end{align}
where we have used $H_{TB}|j,k\rangle=\hbar\varepsilon_{j,k}|j,k\rangle$. The equilibrium density matrix $\rho^0_{jj',kk'}=f_{j,k}\delta_{jj'}\delta_{kk'}$ is constructed by filling electron states according to the Fermi-Dirac distribution occupation numbers $f_{j,k}$ \cite{HL1970}.

We consider excitation of a graphene nanoribbon by continuous-wave (cw) illumination of frequency $\omega$ and in-plane wave vector $q$. An iterative solution of the above equation of motion is now facilitated by expanding the density matrix as
\begin{equation}\label{rhons}
\rho=\sum_{n,s}\rho^{ns}\ee^{\ii s(qmb-\omega t)}.
\end{equation}
In general, the external potential is a function of the position along the nanoribbon direction, according to
\begin{equation}\label{phi_jk}
\phi_{jj',kk'}=\sum_{l,m}a^*_{jl,k}a_{j'l,k'}\phi_l\ee^{\ii\left(q-k+k'\right)mb}/N=\delta_{k-q,k'}\sum_l a^*_{jl,k}a_{j'l,k}\phi_l,
\end{equation}
where
\begin{equation}\label{phi_l}
\phi_{l}=\phi^{\text{ext}}_{l}-2e\sum_{l'}\bar{v}_{ll'}\rho_{l'l'},
\end{equation}
$\phi^{\text{ext}}_l=-\textbf{R}_l\cdot\textbf{E}(t)$ is the external potential describing the incident field $\textbf{E}(t)$, $\bar{v}_{ll'}=\sum_m v_{l0,l'm}$, and $v_{l0,l'm}\ee^{iqmb}$ is the Coulomb interaction between atoms $l$ and $l'$ separated by $m$ unit cells.

Inserting Eqs.\ (\ref{rhons}) and (\ref{phi_jk}) into Eq.\ (\ref{rhojk_eom}), and identifying terms with the same $\ee^{-\ii s\omega t}$ dependence on both sides of the equation, we find
\begin{equation}
\rho^{ns}_{jj',kk'}\ee^{\ii sqmb}=-\frac{e}{\hbar}\sum_{ll'}\frac{\phi^{ns}_l\rho^0_{ll',k-sq,k'}-\phi^{ns}_{l'}\rho^0_{ll',k,k'+sq}}{s\omega+\ii\tau^{-1}/2-\left(\varepsilon_{j,k}-\varepsilon_{j',k'}\right)}a^*_{jl,k}a_{j'l',k'}+\eta^{ns}_{jj',kk'},
\end{equation}
where
\begin{equation}
\eta^{ns}_{jj',kk'}=-\frac{e}{\hbar}\sum^{n-1}_{n'=1}\sum^{n'}_{s'=-n'}\sum_{ll'}\frac{\phi^{n's'}_l\rho^{n-n',s-s'}_{ll',k-s'q,k'}-\phi^{n's'}_{l'}\rho^{n-n',s-s'}_{ll',k,k'+s'q}}{s\omega+\ii\tau^{-1}/2-\left(\varepsilon_{j,k}-\varepsilon_{j',k'}\right)}a^*_{jl,k}a_{j'l',k'}
\end{equation}
and
\begin{equation}
\phi^{ns}_l=\phi^{\text{ext}}_l\delta_{n,1}\left(\delta_{s,-1}+\delta_{s,1}\right)-2e\sum_{l',m}v_{ll',0m}\ee^{\ii sqmb}\rho^{ns}_{l'l'}.
\end{equation}
Moving from $|j,k\rangle$ (state) to $|l,m\rangle$ (site) representation, we obtain the diagonal density matrix elements (within a unit cell of the nanoribbon, \textit{i.e.} for a single $m$) as
\begin{equation}
\rho^{ns}_{ll,q}=\frac{-1}{2e}\sum_{l'}\chi^0_{ll',q}(s\omega)\phi^{ns}_{l'}+\beta^{ns}_{l,q}
\end{equation}
where
\begin{equation}\label{chi0}
\chi^0_{ll',q}(\omega)=\frac{2e^2}{\hbar}\frac{b}{2\pi}\int^{\pi/b}_{-\pi/b}dk\sum_{jj'}\left(f_{j',k-sq}-f_{j,k}\right)\frac{a_{jl,k}a^*_{j'l,k-sq}a^*_{jl',k}a_{j'l',k-sq}}{\omega+\ii\tau^{-1}/2-\left(\varepsilon_{j,k}-\varepsilon_{j',k-sq}\right)}\ee^{-\ii sqmb}
\end{equation}
is the noninteracting RPA susceptibility, and
\begin{equation}
\beta^{ns}_{l,q}=\sum_{jj',kk'}a_{jl,k}a_{j'l,k'}\eta^{ns}_{jj',kk'}\ee^{-\ii sqmb}.
\end{equation}

Here we focus on normally incident light relative to the graphene plane, and thus the external potential does not depend on the position along the nanoribbon direction, \textit{i.e.}, we take $q=0$ in the above expressions. Following a procedure described elsewhere \cite{paper247}, we compute the density matrices $\rho^{ns}_{ll',0}$, from which we obtain the $n$th-order induced charge at site $l$ oscillating with harmonic $s$ as $\rho^{\text{ind}}_l=-2e\rho^{ns}_{ll,0}$. The polarizability per unit length at order $n$ and harmonic $s$ is then given by
\begin{equation}
\alpha^{(n)}_{s\omega}=-\frac{2e}{(E_0)^s}\sum_l\rho^{ns}_{ll,0}\textbf{R}_l\cdot\hat{\textbf{e}}.
\end{equation}
Using this formalism, we compute the linear polarizability $\alpha^{(1)}_{\omega}$, as well as the nonlinear polarizabilities corresponding to SHG $\alpha^{(2)}_{2\omega}$, THG $\alpha^{(3)}_{3\omega}$, and the optical Kerr nonlinearity $\alpha^{(3)}_{\omega}$.

For finite graphene nanoislands, the linear and nonlinear optical response is simulated following the procedure previously described for the analysis of SHG and THG \cite{paper247}, employing a fast Fourier transform (FFT) method to expedite the calculation of the noninteracting RPA susceptibilities \cite{paper183} (see SI for details on the convergence of the FFT method with direct calculations).

\section{Classical electrostatic theory for nanostructured graphene}

We consider a graphene nanostructure with a characteristic size $D$, corresponding to the width of a nanoribbon or the side length of an equilateral triangle that is much less than the wavelength of the incident illumination, for which the electric field is given by $\Eb^{\text{ext}}=E_0\hat{\textbf{e}}\left(\ee^{-\ii\omega t}+\text{c.c.}\right)$, where $\hat{\textbf{e}}$ is the polarization unit vector. We quantify the linear and nonlinear optical response of a finite nanostructure by the $n^{\text{th}}$-order dipole induced along $\hat{\textbf{e}}$ oscillating at harmonic $s$ of the excitation frequency $\omega$, for which the polarizability is given by
\begin{equation}\label{alpha_ns}
\alpha^{(n)}_{s\omega}(\omega)=\frac{1}{(E_0)^n}\int d^2\Rb\left(\hat{\textbf{e}}\cdot\Rb\right)\rho^{\text{ind},(n)}_{s\omega}(\Rb,\omega),
\end{equation}
where the induced charge density is obtained from the surface currents $\jb^{(n)}_{s\omega}(\Rb,\omega)$ using the continuity equation,
\begin{equation}\label{cont_eq}
\rho^{\text{ind},(n)}_{s\omega}(\Rb,\omega)=-\frac{\ii}{s\omega}\nabla_{\Rb}\cdot\jb^{(n)}_{s\omega}(\Rb,\omega),
\end{equation}
and $\Rb=(x,y)$ are 2-D coordinate vectors in the $x$-$y$ plane.

For the linear optical response (taking $n=s=1$ in the above expressions), the surface current is $\jb^{(1)}_{\omega}(\Rb,\omega)=\sigma^{(1)}_{\omega}(\Rb,\omega)\Eb(\Rb,\omega)$, where $\sigma^{(1)}_{\omega}(\Rb,\omega)$ is the linear conductivity of graphene (see Appendix) and $\Eb(\Rb,\omega)=-\nabla_{\Rb}\phi(\Rb,\omega)$ is the total electric field acting on the graphene nanostructure due to the self-consistent potential $\phi$, which is given in the electrostatic approximation by \cite{paper228}
\begin{equation}
\phi(\Rb,\omega)=\phi^{\text{ext}}(\Rb,\omega)+\frac{\ii}{\omega}\int\frac{d^2\Rb'}{|\Rb-\Rb'|}\nabla_{\Rb'}\cdot\sigma^{(1)}_{\omega}(\Rb',\omega)\nabla_{\Rb'}\phi(\Rb',\omega).
\end{equation}
Following the method of Ref.\ \cite{paper228}, we assume that the linear conductivity can be separated as $\sigma^{(1)}_{\omega}(\Rb,\omega)=f(\Rb)\sigma^{(1)}_\omega(\omega)$, where the occupation factor $f(\Rb)=1$ within the graphene structure and is zero everywhere else, and we express Eq.\ (\ref{cont_eq}) in terms of a reduced 2-D coordinate vector $\thetav=\Rb/D$ as
\begin{equation}\label{rho1}
\rho^{\text{ind},(1)}_\omega(\thetav,\omega)=-\frac{\eta^{(1)}_\omega(\omega)}{D}\nabla_{\thetav}\cdot\sqrt{f(\thetav)}\,\varepsv(\thetav,\omega).
\end{equation}
In obtaining the above expression we have defined $\eta^{(n)}_{s\omega}(\omega)=\ii\sigma^{(n)}_{s\omega}(\omega)/s\omega D$ and introduced the normalized electric field $\varepsv(\thetav,\omega)=-\sqrt{f(\thetav)}\nabla_{\thetav}\phi(\thetav,\omega)$, which is expanded in a complete set of eigenmodes $\varepsv_j$ with real eigenvalues $1/\eta_j$ as \cite{paper228}
\begin{equation}\label{E_norm}
\varepsv(\thetav,\omega)=\sum_j\frac{c_j}{1-\eta^{(1)}_\omega(\omega)/\eta_j}\varepsv_j(\thetav),
\end{equation}
where the expansion coefficients are
\begin{equation}
c_j=\int d^2\thetav \varepsv_j(\thetav)\cdot\varepsv\ ^{\text{ext}}(\thetav,\omega)=DE_0\hat{\textbf{e}}\cdot\int d^2\thetav \sqrt{f(\thetav)} \varepsv_j(\thetav),
\end{equation}
and the eigenmodes are orthogonal, i.e.,
\begin{equation}\label{ortho}
\varepsv_j(\thetav)=\int d^2\thetav \varepsv_j(\thetav)\cdot\varepsv_j(\thetav)=\delta_{jj'}.
\end{equation}
Now, assuming that the optical response is dominated by the lowest-order dipolar mode (the $j=1$ term in Eq.\ (\ref{E_norm})), the induced charge density given in Eq.\ (\ref{rho1}) becomes
\begin{equation}\label{rho_ind}
\rho^{\text{ind},(1)}_\omega(\thetav,\omega)=-\frac{\eta^{(1)}_\omega(\omega)}{D}\frac{c_1}{1-\eta^{(1)}_\omega(\omega)/\eta_1}\nabla_{\thetav}\cdot\sqrt{f(\thetav)}\varepsv_1(\thetav).
\end{equation}
We then express Eq.\ (\ref{alpha_ns}) in terms of normalized coordinates $\thetav$ and use Eq.\ (\ref{rho_ind}) to write the linear polarizability as
\begin{equation}
\alpha^{(1)}_{\omega,i}(\omega)=\frac{\eta^{(1)}_\omega(\omega)\xi_1^2 D^3}{1-\eta^{(1)}_\omega(\omega)/\eta_1},
\end{equation}
where we have defined
\begin{equation}
\xi_1=\hat{\textbf{e}}\cdot\int d^2\thetav \thetav \nabla_{\thetav} \cdot \sqrt{f(\thetav)} \varepsv_1(\thetav)=-\hat{\textbf{e}}\cdot\int d^2\thetav \sqrt{f(\thetav)} \varepsv_1(\thetav).
\end{equation}
Moving back to $\Rb$ space, we write $\xi_1=-\hat{\textbf{e}}\cdot\int_S d^2\Rb\,\varepsv_1(\Rb)/D^2$, as given above in the Appendix, by using the function $f(\Rb)$ to restrict the integration to the surface of the graphene nanostructure, $S$, while Eq.\ (\ref{ortho}) guarantees that $\int_S d^2\Rb |\varepsv_1(\Rb)|^2=D^2$.

To describe the nonlinear response associated with second-harmonic generation, we express the second-harmonic current as \cite{paper259}
\begin{equation}
j^{(2)}_{2\omega,i}(\Rb,\omega)=\sigma^{(2)}_{2\omega}(\Rb,\omega)\sum_{jkl}{\bf{\Delta}}^{(2)}_{ijkl}E_j(\Rb,\omega) \partial_k E_l(\Rb,\omega),
\end{equation}
where we assume that the nonlinear conductivity can be written as $\sigma^{(2)}_{2\omega}(\Rb,\omega)=f(\Rb)\sigma^{(2)}_{2\omega}(\omega)$, and we have isolated its tensorial part, ${\bf{\Delta}}^{(2)}_{ijkl}=5\delta_{ij}\delta_{kl}/3-\delta_{ik}\delta_{jl}+\delta_{il}\delta_{jk}/3$. We then express Eq.\ (\ref{cont_eq}), for $n=s=2$, in reduced coordinates as
\begin{equation}
\rho^{\text{ind},(2)}_{2\omega}(\thetav,\omega)=-\frac{\eta^{(2)}_{2\omega}(\omega)}{D^3}\sum_{ijkl}\partial_{\thetav,i} \left[{\bf{\Delta}}^{(2)}_{ijkl}\sqrt{f(\thetav)} \varepsilon_j(\thetav,\omega) \partial_{\thetav,k} \frac{1}{\sqrt{f(\thetav)}}\varepsilon_l(\thetav, \omega)\right].
\end{equation}
Keeping the total electric field to linear order, we again use only the $j=1$ term in Eq.\ (\ref{E_norm}) and write
\begin{equation}
\rho^{\text{ind},(2)}_{2\omega}(\thetav,\omega)=-\frac{\eta^{(2)}_{2\omega}(\omega)}{D^3}\frac{c_1^2}{\left[1-\eta^{(1)}_\omega(\omega)/\eta_1\right]^2}\sum_{ijkl}\partial_{\thetav,i} \left[{\bf{\Delta}}^{(2)}_{ijkl}\sqrt{f(\thetav)} \varepsilon_{1,j}(\thetav) \partial_{\thetav,k} \frac{1}{\sqrt{f(\thetav)}}\varepsilon_{1,l}(\thetav)\right].
\end{equation}
Using the above expression in Eq.\ (\ref{alpha_ns}), expressed in terms of $\thetav$, then yields the nonlinear polarizability for second-harmonic generation,
\begin{equation}
\alpha^{(2)}_{2\omega}(\omega)=\frac{\eta^{(2)}_{2\omega}\xi_1^2\zeta^{(2)}_{2\omega}D^2}{\left[1-\eta^{(1)}_\omega(\omega)/\eta_1\right]^2},
\end{equation}
where we introduce the unitless parameter
\begin{equation}
\zeta^{(2)}_{2\omega}=-\int d^2\thetav\left(\thetav\cdot\hat{\textbf{e}}\right)\sum_{ijkl}\partial_{\thetav,i} \left[{\bf{\Delta}}^{(2)}_{ijkl}\sqrt{f(\thetav)} \varepsilon_{1,j}(\thetav) \partial_{\thetav,k} \frac{\varepsilon_{1,l}(\thetav)}{\sqrt{f(\thetav)}}\right]= \sum_{ijkl}\hat{\textbf{e}}_i{\bf{\Delta}}^{(2)}_{ijkl}\int d^2\thetav\sqrt{f(\thetav)} \varepsilon_{1,j}(\thetav) \partial_{\thetav,k} \frac{\varepsilon_{1,l}(\thetav)}{\sqrt{f(\thetav)}}.
\end{equation}
Note that $\varepsilon_{1,i}(\thetav)/\sqrt{f(\thetav)}$ is proportional to the physical electric field associated with the dipolar plasmon mode, and therefore is continuous across the graphene edge, even in the limit where $f(\thetav)$ jumps from 1 to 0 at the edge itself; the leading factor of $\sqrt{f(\thetav)}$ then limits the integration in the above expression to the region occupied by the graphene, $S$. By moving to $\Rb$ space, and considering (without loss of generality) the response along the $x$-direction, i.e., for $\hat{\textbf{e}}_x$, we may express $\zeta^{(2)}_{2\omega}$ in the form
\begin{equation}
\zeta^{(2)}_{2\omega}=\frac{1}{D}\int_S d^2\Rb \left\{\varepsilon_{1,x}(\Rb) \left[\frac{\partial}{\partial x}\varepsilon_{1,x}(\Rb)+\frac{5}{3}\frac{\partial}{\partial y}\varepsilon_{1,y}(\Rb)\right]+\varepsilon_{1,y}(\Rb)\left[\frac{1}{3}\frac{\partial}{\partial y}\varepsilon_{1,x}(\Rb)-\frac{\partial}{\partial x}\varepsilon_{1,y}(\Rb)\right]\right\},
\end{equation}
which corresponds to the expression provided in the Appendix.

The extension to third-order nonlinearities follows straightforwardly: For third-harmonic generation, we start with the third-order surface current
\begin{equation}
j^{(3)}_{3\omega,i}(\Rb,\omega)=\sigma^{(3)}_{3\omega}(\Rb,\omega)\sum_{jkl}{\bf{\Delta}}^{(3)}_{ijkl}E_j(\Rb,\omega)E_k(\Rb,\omega)E_l(\Rb,\omega),
\end{equation}
where the tensor part is ${\bf{\Delta}}^{(3)}_{ijkl}=\left(\delta_{ij}\delta_{kl}+\delta_{ik}\delta_{jl}+\delta_{il}\delta_{jk}\right)/3$. We isolate the spatial dependence of the third-order conductivity according to $\sigma^{(3)}_{3\omega}(\Rb,\omega)=f(\Rb)^2\sigma^{(3)}_{3\omega}(\omega)$, and so Eq.\ (\ref{cont_eq}) (for $n=s=3$) becomes
\begin{equation}
\rho^{\text{ind},(3)}_{3\omega}(\thetav,\omega)=-\frac{\eta^{(3)}_{3\omega}(\omega)}{D^3}\sum_{ijkl}\partial_{\thetav,i}\left[{\bf{\Delta}}^{(3)}_{ijkl} \sqrt{f(\thetav)} \varepsilon_j(\thetav,\omega)\varepsilon_k(\thetav,\omega) \varepsilon_l(\thetav, \omega)\right].
\end{equation}
Using Eq.\ (\ref{E_norm}), we find that
\begin{equation}
\rho^{\text{ind},(3)}_{3\omega}(\thetav,\omega)=-\frac{\eta^{(3)}_{3\omega}(\omega)}{D^3}\frac{c_1^3}{\left[1-\eta^{(1)}_\omega(\omega)/\eta_1\right]^3}\sum_{ijkl}\partial_{\thetav,i}\left[{\bf{\Delta}}^{(3)}_{ijkl} \sqrt{f(\thetav)} \varepsilon_{1,j}(\thetav) \varepsilon_{1,k}(\thetav) \varepsilon_{1,l}(\thetav)\right],
\end{equation}
from which Eq.\ (\ref{alpha_ns}) provides the nonlinear polarizability corresponding to third-harmonic generation,
\begin{equation}
\alpha^{(3)}_{3\omega}(\omega)=\frac{\eta^{(3)}_{3\omega}\xi_1^3\zeta^{(3)}_{3\omega}D^3}{\left[1-\beta^{(1)}_\omega(\omega)/\eta_1\right]^3},
\end{equation}
where
\begin{equation}
\zeta^{(3)}_{3\omega}=\int d^2\thetav \left(\thetav\cdot\hat{\textbf{e}}\right) \sum_{ijkl} \partial_{\thetav,i}\left[ {\bf{\Delta}}^{(3)}_{ijkl} \sqrt{f(\thetav)} \varepsilon_{1,j}(\thetav) \varepsilon_{1,k}(\thetav) \varepsilon_{1,l}(\thetav)\right]=-\sum_{ijkl}\hat{\textbf{e}}_i {\bf{\Delta}}^{(3)}_{ijkl} \int d^2\thetav \sqrt{f(\thetav)} \varepsilon_{1,j}(\thetav) \varepsilon_{1,k}(\thetav) \varepsilon_{1,l}(\thetav).
\end{equation}
The above expression for the unitless parameter $\zeta^{(3)}_{3\omega}$ can be simplified by moving to $\Rb$ coordinates and using the fact that $f(\Rb)=1$ for $\Rb\in S$, and is zero otherwise, to write
\begin{equation}
\zeta^{(3)}_{3\omega}=-\left(1/D^2\right)\hat{\textbf{e}}\cdot\int_S d^2\Rb\,\varepsv_1(\Rb) \cdot \varepsv_1(\Rb)\,\varepsv_1(\Rb), 
\end{equation}
corresponding to the expression provided in the Appendix.

In an analogous manner, the nonlinear polarizability corresponding to the Kerr nonlinearity (Eq.\ (\ref{alpha_ns}) for $n=3$, $s=1$) is obtained from the third-order surface current
\begin{equation}
j^{(3)}_{\omega,i}(\Rb,\omega)=\sigma^{(3)}_{3\omega}(\Rb,\omega)\sum_{jkl}{\bf{\Delta}}^{(3)}_{ijkl}E_j(\Rb,\omega)E^*_k(\Rb,\omega)E_l(\Rb,\omega)
\end{equation}
as
\begin{equation}
\alpha^{(3)}_{\omega}(\omega)=\frac{\eta^{(3)}_{\omega}(\omega)\xi_1^3\zeta^{(3)}_\omega D^3}{\left|1-\eta^{(1)}_\omega(\omega)/\eta_1\right|^2\left[1-\eta^{(1)}_\omega(\omega)/\eta_1\right]}
\end{equation}
where
\begin{equation}
\zeta^{(3)}_{\omega}=\int d^2\thetav \left(\thetav\cdot\hat{\textbf{e}}\right) \sum_{ijkl}\partial_{\thetav,i} \left[ {\bf{\Delta}}^{(3)}_{ijkl} \sqrt{f(\thetav)} \varepsilon_{1,j}(\thetav)\varepsilon^*_{1,k}(\thetav)\varepsilon_{1,l}(\thetav)\right]=-\sum_{ijkl}\hat{\textbf{e}}_i{\bf{\Delta}}^{(3)}_{ijkl}\int d^2\thetav \sqrt{f(\thetav)}\varepsilon_{1,j}(\thetav)\varepsilon^*_{1,k}(\thetav)\varepsilon_{1,l}(\thetav).
\end{equation}
After performing the summation in the above expression, we obtain
\begin{equation}
\zeta^{(3)}_{\omega}=-\frac{1}{D^2}\hat{\textbf{e}}\cdot\int_S d^2\Rb\left\{ \frac{2}{3}\left|\varepsv_1(\Rb)\right|^2 \varepsv_1(\Rb)+\frac{1}{3}\varepsv_1(\Rb)\cdot\varepsv_1(\Rb)\left[\varepsv_1(\Rb)\right]^*\right\},
\end{equation}
which coincides with the expression provided in the Appendix.

%%%%%%%%%%%%%%%%%%%%%%%%%%%%%%%%%%%%%%%%%%%%%%%%%%%%%%%%%%%%%%%%%%%%%%%%%%

\section{Temperature-dependence of the nonlinear response}

Through iterative solution of the Boltzmann transport equation (in the local limit),
\begin{equation}\label{boltz}
\frac{\partial f_\kb(\textbf{r}, t)}{\partial t}-\frac{e}{\hbar}\textbf{E}\cdot\nabla_\kb f_\kb(\textbf{r}, t)=-\frac{1}{\tau}\left[ f_\kb(\textbf{r}, t) -  f_\kb^0(\varepsilon_k)\right],
\end{equation}
we take $\Eb(t)=\Eb_{\omega}(\ee^{-\ii\omega t}+\ee^{\ii\omega t})$ and find that, to first-order,
\begin{equation}\label{sigma_1}
\sigma^{(1)}_\omega(\omega)=\ii\frac{e^2}{\pi\hbar^2}\frac{F^{(1)}_\kb}{\omega+\ii\tau^{-1}},
\end{equation}
with
\begin{equation}
F^{(1)}_\kb=\int^{\infty}_{-\infty}d\varepsilon_k\left|\varepsilon_k\right|\frac{\partial f^{0}_\kb}{\partial\varepsilon_k}=E_F+2k_BT\log\left(1+\ee^{-E_F/k_BT}\right) \nonumber,
\end{equation}
so that Eq.\ (\ref{sigma_1}) leads to Eq.\ (5) in the $T\rightarrow0$ limit. Actually, the linear result is virtually unchanged by considering $T=300$\,K, for which we find $F^{(1)}_\kb\simeq E_F$ for the Fermi energies considered in this work, \textit{i.e.} for $0.2\leq E_F/\text{eV}\leq2.0$.
At third-order, we obtain expressions for the local intraband conductivity of graphene as
\begin{equation}
\sigma^{(3)}_{s\omega}(\omega)=-\ii\frac{e^4v_F^2}{\pi^2\hbar^2}\frac{-sF^{(3)}_\kb}{D_{s\omega}(\omega)},
\end{equation}
where
\begin{equation}\label{Fk3}
F^{(3)}_\kb=\int^\infty_{-\infty}d\varepsilon_k\left(\frac{-1}{|\varepsilon_k|}\frac{\partial f^0_\kb}{\partial\varepsilon_k}+\frac{\partial^2 f^0_\kb}{\partial\varepsilon_k^2}+\left|\varepsilon_k\right|\frac{\partial^3 f^0_\kb}{\partial\varepsilon_k^3}\right),
\end{equation}
and $D_{3\omega}(\omega)=(\omega+\ii\tau^{-1})(2\omega+\ii\tau^{-1})(3\omega+\ii\tau^{-1})$ for third-harmonic generation ($s=3$) or $D_{\omega}(\omega)=(\omega+\ii\tau^{-1})(2\omega+\ii\tau^{-1})(-\omega+\ii\tau^{-1})$ for the Kerr nonlinearity ($s=1$).
Now, at zero temperature, we have $\partial f^0_\kb/\partial\varepsilon_k=-\delta\left(E_F-\varepsilon_k\right)$, and Eq.\ (\ref{Fk3}) reduces to $F_\kb=-3\pi/4$, after having used the general relation
\begin{equation}
\int dx \frac{\partial^n f(x)}{\partial x^n}\delta(a-x)=(-1)^n\left.\frac{\partial^n f(x)}{\partial x^n}\right|_{x=a}.
\end{equation}
Thus, we recover
\begin{equation}
\sigma^{(3)}_{3\omega}(\omega)=\frac{3\ii e^4 v_F^2}{4\pi\hbar^2 E_F}\frac{1}{ (\omega+\ii\tau^{-1})(2\omega+\ii\tau^{-1})(3\omega+\ii\tau^{-1})}
\end{equation}
and
\begin{equation}
\sigma^{(3)}_{\omega}(\omega)=\frac{9\ii e^4 v_F^2}{4\pi\hbar^2 E_F}\frac{1}{(\omega+\ii\tau^{-1})(-\omega+\ii\tau^{-1})(2\omega+\ii\tau^{-1})},\nonumber
\end{equation}
corresponding to Eqs.\ (9) and (10) in the Appendix, and in agreement with the result of Ref.\ \cite{PBS14}. These third-order expressions differ in multiplicative factors from those reported in a previous study \cite{M15_2}, where the phenomenological decay is introduced in the Fourier integrals; if in Eq.\ (\ref{boltz}) we had instead defined $\Eb(t)=\int^{\infty}_{-\infty}d\omega\Eb_{\omega}\ee^{-i\omega t+t/\tau}$, we would recover the purely intraband contributions to the third-order currents obtained from Ref.\ \cite{M15_2}:
\begin{equation}
\sigma^{(3)}_{3\omega}(\omega)=\frac{\ii e^4 v_F^2}{8\pi\hbar^2 E_F}\frac{1}{(\omega+\ii\tau^{-1})^3}
\end{equation}
and
\begin{equation}
\sigma^{(3)}_{\omega}(\omega)=\frac{3\ii e^4 v_F^2}{8\pi\hbar^2 E_F}\frac{1}{(\omega+\ii\tau^{-1})^2(-\omega+\ii\tau^{-1})}.\nonumber
\end{equation}

For the intraband third-order nonlinear conductivities at non-zero temperatures, we find that Eq.\ (\ref{Fk3}) reduces to
\begin{equation}
F^{(3)}_\kb=\frac{-3\pi}{4k_BT}\left\{\int^\infty_{-\infty}\frac{d\varepsilon_k}{|\varepsilon_k|}\frac{\ee^{\left(\varepsilon_k-E_F\right)/k_BT}}{\left[\ee^{\left(\varepsilon_k-E_F\right)/k_BT}+1\right]^2}+\frac{2\ee^{-E_F/k_BT}}{(\ee^{-E_F/k_BT}+1)^2}\right\}.
\end{equation}
Note that the integrand in the first term above has a singularity at $\varepsilon_k=0$, and so $F^{(3)}_\kb$ diverges. This suggests that the perturbation theory (up to third order) is inadequate for dealing with finite temperatures, in the same way it fails as $E_F\rightarrow0$ (consider the $1/E_F$ dependence of $\sigma^{(3)}_{s\omega}$). However, we expect that the $T=0$ description of the third-order, purely intraband conductivities should also describe the $T=300$\,K case reasonably well, considering the negligible change in the linear response (see Eq.\ (\ref{sigma_1})), as well as those in the nonlinear response for graphene nanoribbons predicted by atomistic simulations (see Fig.\ \ref{T0}).
\begin{figure*}[t]
\includegraphics[width=0.75\textwidth]{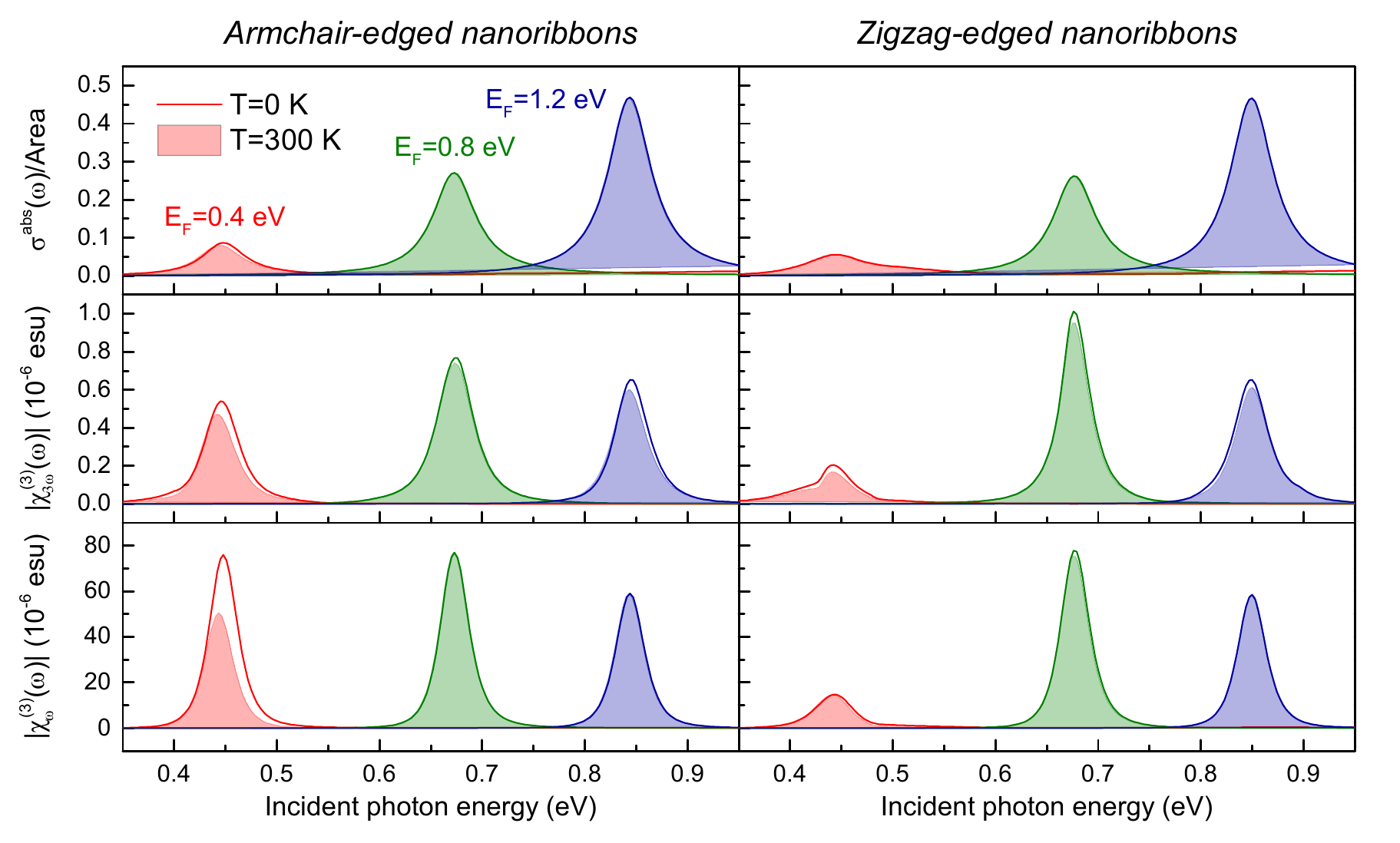}
\caption{\label{T0} {\bf Temperature dependence of the nonlinear optical response in nanoribbons.} We show spectra for the linear absorption cross-section (upper panels), along with the nonlinear polarizabilities corresponding to third-harmonic generation (middle panels), and the Kerr nonlinearity (lower panels) for $\sim10$\,nm armchair- (left panels) and zigzag-edged (right panels) nanoribbons at zero temperature (filled curves) and for $T=300$\,K (regular curves). Different Fermi energies are considered here, as indicated by the color-coded numerical values in the upper-left panel.}
\end{figure*}

%%%%%%%%%%%%%%%%%%%%%%%%%%%%%%%%%%%%%%%%%%%%%%%%%%%%%%%%%%%%%%%%%%%%%%%%%%

\section{Regarding interband contributions to the nonlinear optical response}

Computing the full third-order nonlinear optical conductivity of extended graphene, \textit{i.e.}, including contributions from both intraband and interband optical transitions, is presently an area of active study, with various approaches in the literature that apparently yield different results (see for example the work of Cheng {\it et al.} in Refs.\ \cite{CVS14,CVS15} and by Mikhailov in Ref.\ \cite{M15_2}). In this work, we are primarily interested in the nonlinear optical response of graphene nanostructures enhanced by plasmonic exctations, and thus we restrict our investigation to the regime where only intraband contributions will contribute significantly to the nonlinear conductivities. Incidentally, it has been argued that interband transitions only become important at energies $\hbar\omega\geq2E_F$, and even then their effect is orders of magnitude smaller than the contribution of intraband transitions \cite{M15_2}.

A relatively simple, straightforward improvement to the classical description of the plasmon-enhanced nonlinear response in nanostructured graphene is to include the effect of interband optical transitions in the linear response, which can be accomplished by using the linear conductivity for graphene obtained from the random-phase approximation in the local limit \cite{paper235},
\begin{equation}\label{sigma_RPA}
\sigma^{(1)}_\omega(\omega)=-\ii\frac{e^2}{\pi\hbar^2}\frac{1}{\omega+\ii\tau^{-1}}\left[F^{(1)}_\kb+\int^{\infty}_{-\infty} d\varepsilon_k \frac{(\varepsilon_k/|\varepsilon_k|)f^0_\kb(\varepsilon_k)}{1-4\varepsilon_k^2/\left[\hbar^2\left(\omega+\ii\tau^{-1}\right)^2\right]}\right].
\end{equation}
The linear conductivity enters the factors $\eta^{(1)}_\omega$ in Eqs.\ (1-4) of the Appendix, effectively describing the local field generated by the graphene nanostructure with greater accuracy. An accurate description of the local field enhancement is arguably more important when describing the nonlinear response at plasmon resonances than to include the effect of interband optical transitions in the nonlinear conductivities. Indeed, as we show in Fig.\ \ref{LRPA300} for graphene nanoribbons, the peak intensities for the linear and nonlinear responses are in better agreement with those predicted in quantum-mechanical simulations when we use Eq.\ (\ref{sigma_RPA}) to describe the linear conductivity in graphene, where the most noticeable improvements appear for resonances above or near the Fermi level.
\begin{figure*}[t]
\includegraphics[width=0.75\textwidth]{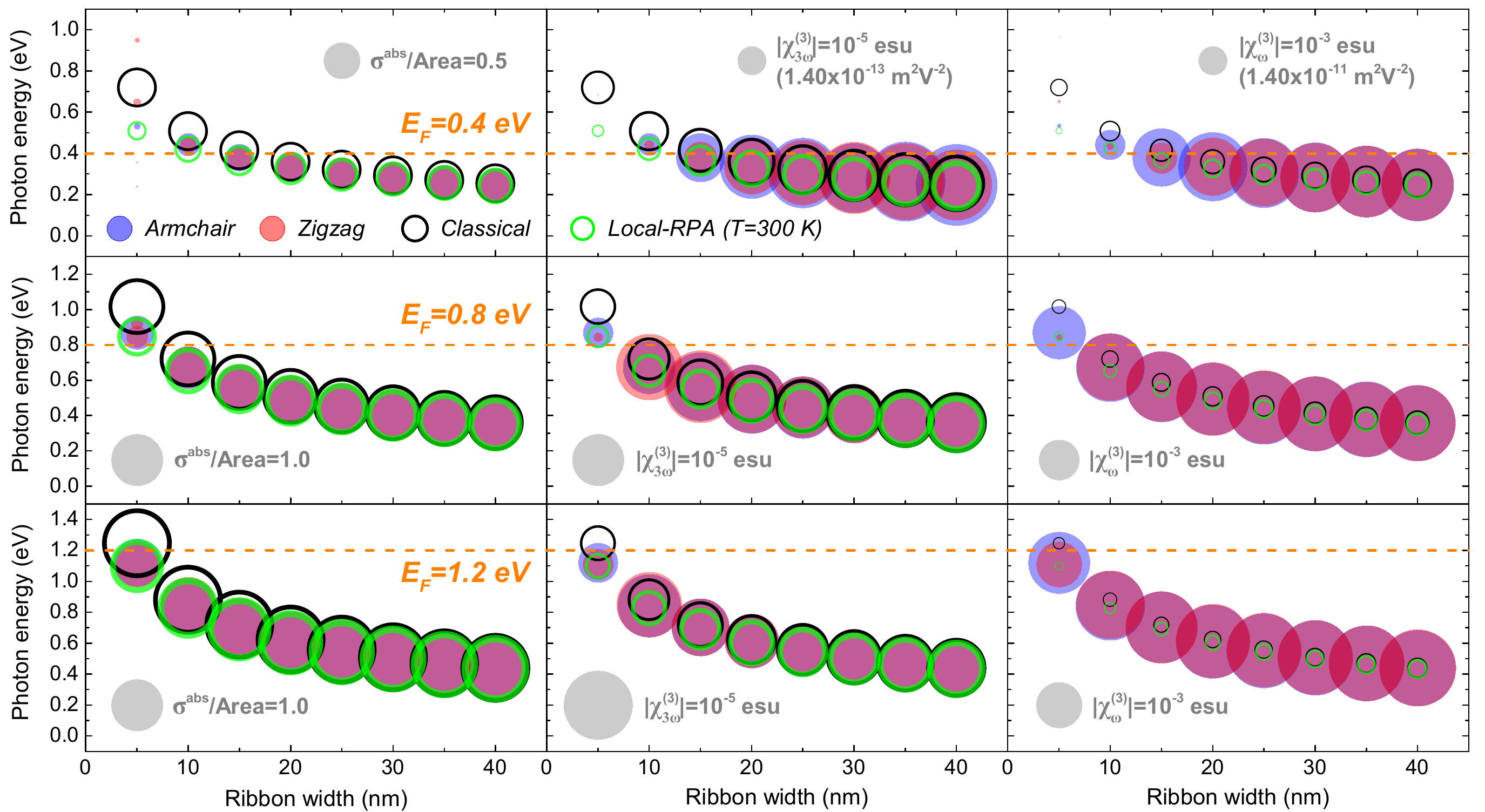}
\caption{\label{LRPA300} {\bf Atomistic description of optical response for nanoribbons compared with classical simulations, including interband transitions in the linear response.} The symbols indicate the peak maxima for the linear absorption cross-section (left column), as well as the nonlinear polarizabilities corresponding to THG (central column) and the Kerr nonlinearity (right column), calculated for graphene nanoribbons illuminated by light polarized perpendicular to their direction of translational symmetry. Results are presented for QM calculations of armchair (solid blue circles) and zigzag (solid red circles) ribbons, compared with classical simulations, where we show results for only intraband contributions in the linear and nonlinear conductivities (open black circles) or with interband contributions included in the linear conductivity (open green circles).}
\end{figure*}

%%%%%%%%%%%%%%%%%%%%%%%%%%%%%%%%%%%%%%%%%%%%%%%%%%%%%%%%%%%%%%%%%%%%%%%%%%

\section{Nonlinear refractive index and nonlinear absorption}

The complex third-order susceptibility oscillating at the fundamental frequency of illumination, $\chi^{(3)}_{\omega}$, which we have referred to here as the Kerr nonlinearity, is related to the nonlinear refractive index $n_2$ and nonlinear absorption coefficient $\beta$ of a medium. These quantities are typically what are measured directly in nonlinear optical experiments, from which values for the nonlinear susceptibility are inferred. The relations among a medium's $\chi^{(3)}_{\omega}$, $n_2$, and $\beta$ are provided in Ref.\ \cite{CS04} as
\begin{equation}
n_2=\frac{3}{4\epsilon_0c\left(n_0^2+k_0^2\right)}\left(\text{Re}\{\chi^{(3)}_\omega\}+\frac{k_0}{n_0}\text{Im}\{\chi^{(3)}_\omega\}\right)
\end{equation}
and
\begin{equation}
\beta=\frac{3}{2\epsilon_0c^2\left(n_0^2+k_0^2\right)}\left(\text{Im}\{\chi^{(3)}_\omega\}-\frac{k_0}{n_0}\text{Re}\{\chi^{(3)}_\omega\}\right),
\end{equation}
where $n_0=\text{Re}\{(1+\chi^{(1)}_\omega)^{1/2}\}$ and $k_0=\text{Im}\{(1+\chi^{(1)}_\omega)^{1/2}\}$, with all of the above quantities in SI units. The above expressions are intended to properly account for the interplay between the complex first- and third-order susceptibilities in highly-absorbing media \cite{CS04}, and we use them here to show the frequency-dependence of $n_2$ and $\beta$ for the $\sim10$\,nm graphene nanoribbons considered in Fig.\ 1.
\begin{figure*}[t]
\includegraphics[width=0.75\textwidth]{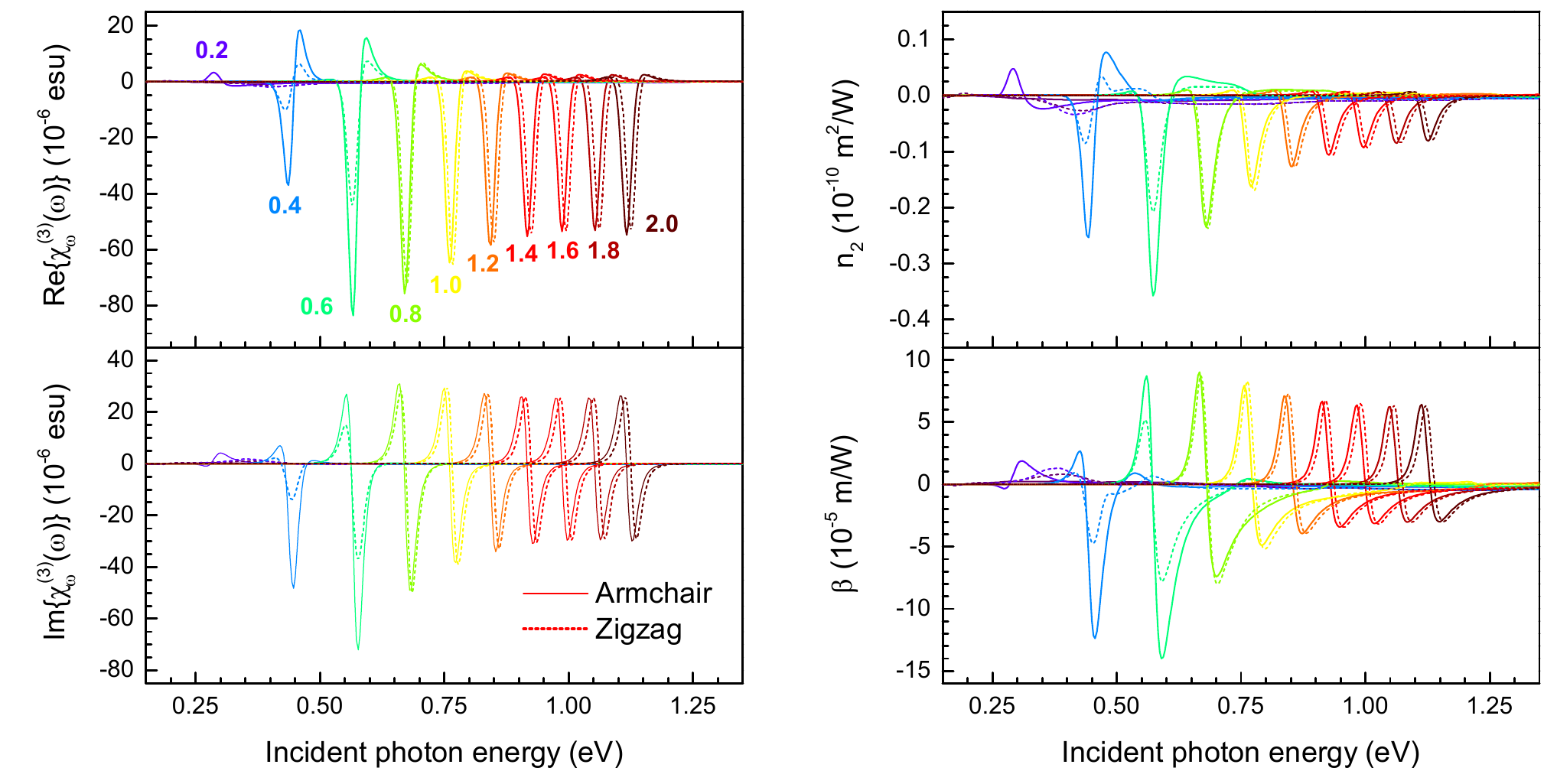}
\caption{{\bf Kerr-type nonlinearities in graphene nanoribbons} We present the real and imaginary parts of the third-order nonlinear susceptibilities used to produce Fig.\ 1d (left column, in electrostatic units), along with the corresponding values of the nonlinear refractive index, $n_2$, and the nonlinear absorption coefficient, $\beta$ (right column, in SI units). Different Fermi energies (color-coded numerical values in (a), eV) are considered, and the ribbon width is $\approx10\,$nm.}
\end{figure*}

%%%%%%%%%%%%%%%%%%%%%%%%%%%%%%%%%%%%%%%%%%%%%%%%%%%%%%%%%%%%%%%%%%%%%%%%%%

\section{Convergence of the fast-Fourier transform method for nanoislands}

In this work, the nonlinear response for finite graphene nanoislands is described using the perturbative expansion procedure for the single-electron density matrix outlined in Ref.\ \cite{paper247}. In this approach, we compute several noninteracting RPA susceptibilities, $\chi^0_{ll'}(s\omega)$, which have the same form as Eq.\ (23) of the Appendix (but lack any momentum dependence), to obtain the nonlinear polarizabilities for a single nanoisland. To expedite these computations, which require summing $\sim N^4$ terms, $N$ being the number of carbon atoms in a nanoisland, we employ a fast-Fourier transform (FFT) method, as described in Ref.\ \cite{paper183}, for which only $\sim N^3$ operations are required. To perform the FFT, a finite grid of $N_\omega$ equally-spaced frequencies must be defined, and it is this discretization that determines the convergence of the $\chi^0_{ll'}(s\omega)$ computed using the FFT method with that obtained from a direct evaluation. In Fig.\ \ref{FFT}, we show that while the FFT method can provide excellent convergence for $N_\omega=10^{12}$ frequencies, particularly in the linear response, more satisfactory results for the nonlinear polarizabilities are obtained with $10^{13}$ frequencies. Note that while this convergence is rather independent of a graphene nanoisland's size, more frequencies are required as the phenomenological relaxation rate is reduced.
\begin{figure*}[t]
\includegraphics[width=0.75\textwidth]{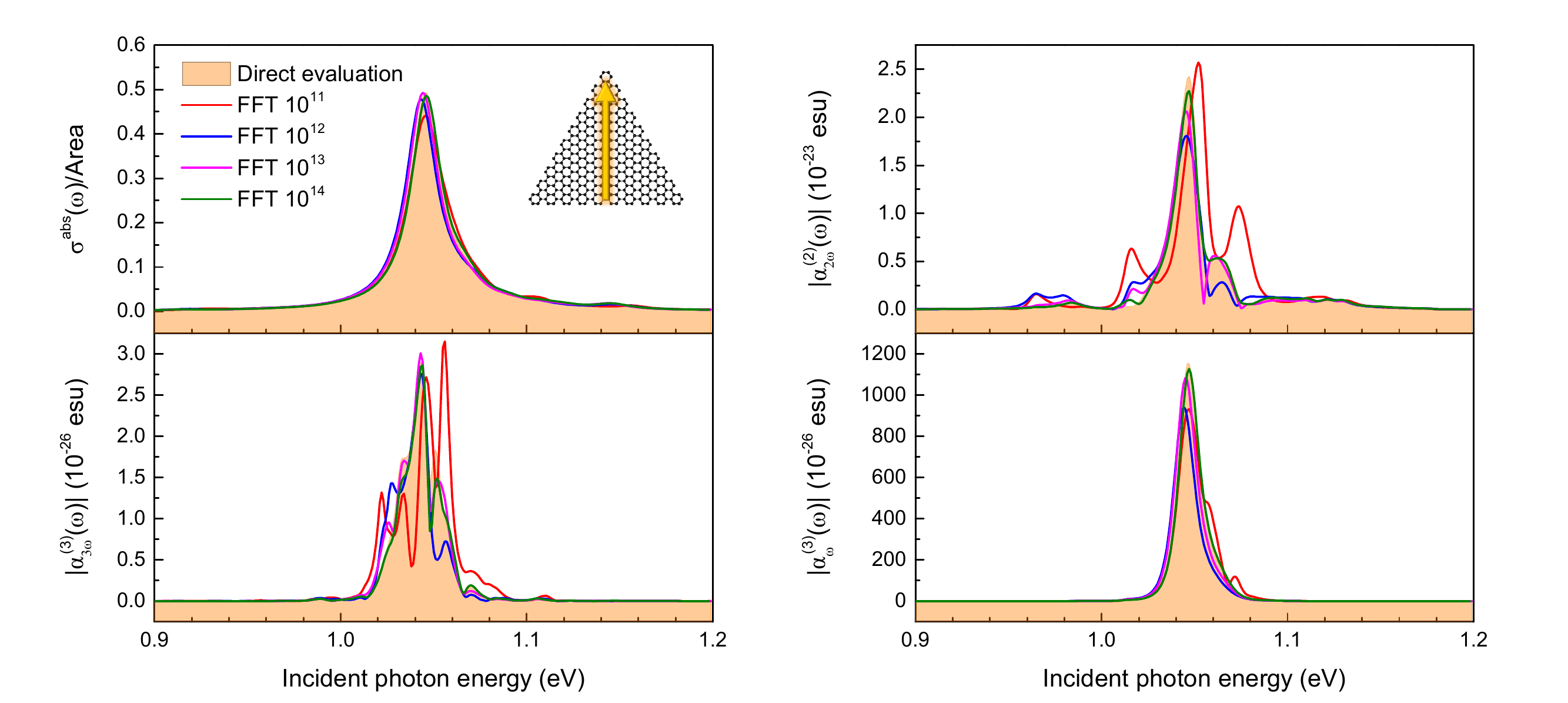}
\caption{\label{FFT} {\bf Convergence of FFT method for graphene nanoislands.} We show spectra for the linear absorption cross-section (upper left), along with the nonlinear polarizabilities corresponding to second-harmonic generation (upper right), third-harmonic generation (lower left), and the Kerr nonlinearity (lower right) for a $\sim4.4$\,nm, armchair-edged nanotriangle, with light polarized perpendicular to one of the triangle sides. Results are shown for a simulation in which the RPA susceptibility is evaluated directly (filled curves) along with those computed using the fast-Fourier transform (FFT) method, using the indicated number of frequencies in the Fourier integral.}
\end{figure*}

\end{widetext}

\acknowledgments

This work has been supported in part by the Spanish MINECO (MAT2014-59096-P and SEV-2015-0522) and the European Commission (Graphene Flagship CNECT-ICT-604391 and FP7-ICT-2013-613024-GRASP). The authors thank Andrea Marini and Renwen Yu for useful discussions.

%\bibliographystyle{apsrev}
%\bibliography{../../../bibtex/refs}

\end{document}